\documentclass[twocolumn,superscriptaddress,prl]{revtex4}
\usepackage{color}
\usepackage{graphicx}
\usepackage{epsfig}
\usepackage{amsmath, amssymb}
\usepackage{verbatim}
\usepackage{bm}
\usepackage{mathrsfs}

\newcommand{\gammabar}{\ensuremath\gamma\kern-0.53em-}

\begin{document}

\title{Experimental Proposal to Detect Topological Ground State Degeneracy}
\author{Maissam Barkeshli}
\affiliation{Station Q, Microsoft Research, Santa Barbara, California 93106, USA}
\author{Yuval Oreg}
\affiliation{Department of Condensed Matter Physics, Weizmann Institute of Science, Rehovot, 76100, Israel}
\author{Xiao-Liang Qi}
\affiliation{Department of Physics, Stanford University, Stanford, California 94305, USA }

\begin{abstract}
One of the most profound features of topologically ordered states of matter, such as the fractional quantum Hall (FQH)
states, is that they possess topology-dependent ground state degeneracies that are robust to all local perturbations.
Here we propose to directly detect these topological degeneracies in an experimentally
accessible setup. The detection scheme uses electrical conductance
measurements in a double layer FQH system with appropriately patterned top and bottom gates.
We discuss two experimental platforms; in the first, the detection of topologically degenerate states
coincides with the detection of $Z_N$ parafermion zero modes. We map the relevant physics to a
single-channel $Z_N$ quantum impurity model, providing a novel generalization of the Kondo model. Our proposal
can also be adapted to detect the $Z_N$ parafermion zero modes recently discovered in FQH line junctions
proximitized with superconductivity.
\end{abstract}

\maketitle

The discovery of topologically ordered states, such as the fractional quantum Hall (FQH) states,
has been one of the most important developments in modern condensed matter physics \cite{wen04,nayak2008}.
Topologically ordered states provide examples of collective behavior beyond the Ginzburg-Landau paradigm of
symmetry-breaking, and give rise to fractionalized quasiparticle excitations.

\begin{figure}
\includegraphics[width=3.4in]{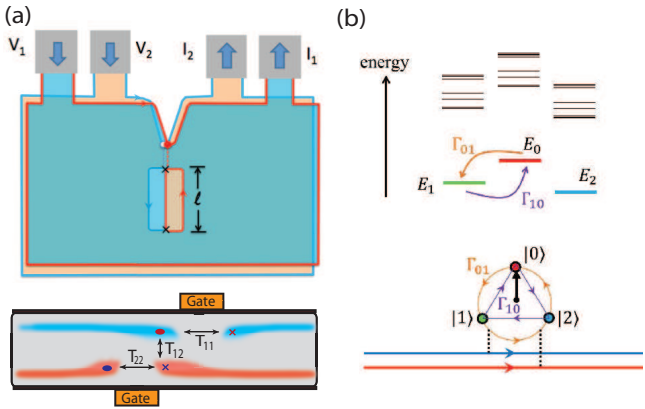}
\caption{\label{setupFig}
(a) Top view and cross-section of proposed experimental setup.
Each layer forms a $1/3$ Laughlin state; top and bottom gates are offset in the lateral direction.
The ends of the gates, marked by crosses, localize $Z_3$ parafermion zero modes with projective non-Abelian braid statistics \cite{barkeshli2013genon}.
(b) (Upper panel) The split topological ground states (green, red and blue levels) and the excitations in each topological sector (black levels) for a finite size topological line junction (TLJ).
(Lower panel) The three topological states of the TLJ can be modelled
by a $Z_3$ quantum clock degree of freedom, illustrated by the three dots. Quasiparticle tunneling
between the two outer edges (blue and red horizontal lines) shifts the state of the $Z_3$
clock in a cyclic manner, illustrated by the arrows around the circle.
}
\end{figure}

A profound feature of topological order is the existence of topology-dependent ground state
degeneracies which are robust to any local perturbations \cite{wen1989,wen1990b}. The $\frac{1}{m}$ Laughlin
FQH states, for example, possess $m^g$ topologically degenerate states on a closed manifold of genus $g$.
While these topology-dependent degeneracies have played an important
role both conceptually and as diagnostics in computational studies for nearly 25 years
and may be used for topological quantum computation \cite{nayak2008}, to date it has not been clear,
even in principle, to what extent their experimental detection is possible.

In this paper, we propose an experiment to directly detect these topological degeneracies in FQH states. Our proposed setup is shown in Fig. \ref{setupFig}, which is referred to as a topological line junction (TLJ) in Ref. \cite{barkeshli2013}. The physical system is a bilayer FQH state, in which each layer is in a $1/m$ Laughlin state, and the filling fraction is the same in the two layers. This is often referred to as an $(mm0)$ state. A pair of top and bottom gates are applied to create two antidots, one in each layer. The two antidots (red and blue rectangles in Fig. \ref{setupFig}) are aligned such that the left edge of the top antidot is right on top of the right edge of the bottom antidot. If the top and bottom gates are sufficiently aligned \footnote{A systematic study of the misalignment error-tolerance is the subject of future work.
An alignment accuracy on the order of the distance between the gates and the 2DEGs is necessary; we expect that $\approx 50 nm$ is within
experimental capability.} and the interlayer distance is suitable, electron backscattering between counterpropagating edge states of \it different \rm layers ($T_{12}$ in Fig. \ref{setupFig}) can open a gap along the edge and effectively build a ``bridge" between two layers, across which topological quasiparticles can tunnel \cite{barkeshli2013}
at low energies. The edge states along the outer edge of the sample are brought near the ends of the TLJ to form a quantum point contact (QPC). The physical quantity of interest is the inter-layer tunneling current induced by this QPC, which can be measured with separate voltage and current probes connecting to each layer. A variation of this setup, using electron-hole bilayers, is discussed in Sec. I of the Supplemental Materials.

In Ref. \cite{barkeshli2013}, it was shown using recent theoretical advances \cite{barkeshli2012a,barkeshli2013genon,barkeshli2010},
that the bilayer $1/m$ Laughlin state with such TLJ's is equivalent to a single layer $1/m$ state on a surface with nontrivial genus, which has topological ground state degeneracies. In addition, the end points of the TLJ become non-Abelian twist defects, referred to as genons, which localize ``parafermion'' zero modes \cite{barkeshli2012a, barkeshli2013genon,lindner2012,clarke2013,cheng2012,fendley2012,barkeshli2013defect,barkeshli2013defect2,clarke2013b} 
(marked by crosses in Fig. \ref{setupFig}). Parafermion zero modes are generalizations of Majorana zero modes \cite{read2000,kitaev2001,fu2008,fu2009d,sau2010,lutchyn2010,oreg2010,alicea2012review}. Ref. \cite{barkeshli2013} has proposed that tunneling into the end points of the TLJ can measure the parafermion zero modes, but the proposed experiment there does not directly measure the topological degeneracy $m$. Directly measuring the topological degeneracy is the goal of this paper.

The key point of the proposal presented here is to consider the small finite-size splitting between the topologically degenerate ground states, which is proportional to $e^{-\ell / \xi_{loc}}$. Here $\ell$ is the antidot size (Fig. \ref{setupFig}) and $\xi_{loc}$ is the localization length of the edge states. 
In the bilayer $(mm0)$ states, the $m$ topologically degenerate ground states can be labeled by the quasiparticle charge at the antidot in the top layer $Q=\frac pm e,~p=0,1,...,m-1$, while the bottom layer has opposite charge $-Q$ so that there is no charging energy. When the splitting is much smaller than the quasiparticle excitation energy of the antidot ($\propto 1/l$), the whole antidot can be considered to be a $m$-state ``quantum clock", and the quasiparticle tunneling at the QPC induces a cyclic permutation of these $m$ states. 

\begin{figure}
\includegraphics[width=3.5in]{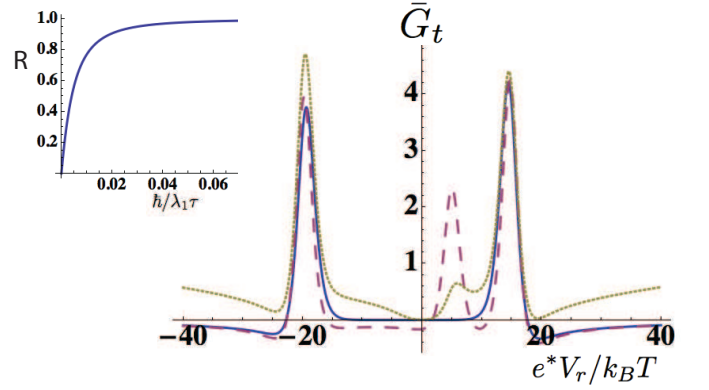}
\caption{\label{condFig} Plot of the normalized DC interlayer tunneling conductance
from the master equation calculation of the $Z_N$ quantum impurity model, where
$\bar{G}_t = \frac{\hbar \Lambda_0 k_B T}{(e^*)^2 |\lambda_1|^2} \frac{d I_t}{d V_r}$. We have chosen
$E_{10}/k_BT = 5$, $E_{21}/k_BT = 15$, $k_BT/\Lambda_0 = 200$, $\lambda_{\pm 1}/\Lambda_0 = 10^{-3}$,
$\lambda_{l} = 0$ for $|l|>2$, and here we have reinstated Boltzmann's constant $k_B$. 
Assuming the ultraviolet cutoff is on the order of the bulk gap
for a typical $1/3$ Laughlin state, $\Lambda_0 \approx 5$ Kelvin, the above corresponds to operating
temperatures of $T = 25 mK$, and topological ground state energy splittings $\sim 100 - 500 mK$.
The different curves are for \{$\lambda_{\pm 2}^2/\lambda_{\pm 1}^2 = 0$, $1/ \tau = 0$\} (blue solid),
\{$\lambda_{\pm 2}^2/\lambda_{\pm 1}^2 = 0$, $\hbar /\tau = 0.005\lambda_1$\} (purple dashed),
\{$\lambda_{\pm 2}^2/\lambda_{\pm 1}^2 = 30$, $1/\tau = 0$\} (yellow dotted).
The last choice of parameters is not expected to be physically relevant, but included here to illustrate
additional features in the clock model due to higher order tunneling terms $\lambda_l$, for $|l| > 1$.
Left inset: ratio of peaks: $R \equiv G_t[E_{10}]/G_t[E_{21}]$, as a function of $\hbar/\lambda_1\tau$, with 
$\lambda_{\pm2} = 0$.
}
\end{figure}

We describe the quantum tunneling problem in this clock model within a master equation approach. In the limit of \it weak \rm quasiparticle tunneling, we show that the energy splittings between the different topological sectors can be probed by resonances in the interlayer conductance, with the number of peaks and their positions being directly related to the number of different topological states and their energy splittings (see Fig. \ref{condFig}). Therefore this experiment directly probes the topological degeneracy in FQH states. Our approach is a generalization of the detection of Majorana zero modes in superconducting wires \cite{dassarma2012} to more generic parafermion zero modes, which can also be applied to other realizations of parafermion zero modes proposed in FQH line junctions proximitized with superconductivity \cite{lindner2012,clarke2013,cheng2012,vaezi2013,mong2013,oreg2013,barkeshli2013defect,barkeshli2013defect2,clarke2013b}.

The rest of the paper is organized as follows. First, we characterize the antidot edge states by a chiral Luttinger liquid theory, based on which we discuss the different phases of the antidot and the condition of realizing a TLJ connecting the two layers. Secondly, we obtain the quantum clock model to describe the tunneling through the QPC. Thirdly, we use the master equation approach to calculate the tunneling conductance in the weak tunneling limit, and discuss the resonance features obtained. An alternative physical setup involving electron-hole bilayers, along with other additional discussions, are included in the Supplemental Materials.

\it Phase diagram of the line junction \rm --  
We first discuss the phase diagram of the line junction formed by the edge states along the antidot.
The edge theory description has been presented in Ref. \cite{barkeshli2013} but the full phase diagram
has not been discussed there. 
In the presence of the gates, part of the FQH fluid is depleted, leading to a pair of counterpropagating
edge states in each layer (see cross-section in Fig. \ref{setupFig} (a)). The Hamiltonian for the edge states
is of the form \cite{wen1992} $H = \int dx (\mathcal{H}_0 + \mathcal{H}_{tun})$, with
$\mathcal{H}_0 = m \pi v_0 \sum_{\alpha,\beta,I,J} n_{\alpha I} \lambda_{IJ}^{\alpha \beta} n_{\beta J}$,
and
\begin{align}
\mathcal{H}_t = \sum_{I,J} |T_{IJ}| \cos(m (\phi_{LI} + \phi_{RJ}) + \theta_{IJ}).
\end{align}
$\mathcal{H}_0$ and $\mathcal{H}_t$ are the Hamiltonian densities of the kinetic/interaction terms and the backscattering
terms, respectively. $\alpha$, $\beta$ = $L, R$ and $I,J = 1,2$ denote the chirality and layer index of
the edge states, respectively. The densities are $n_{\alpha I} = \frac{1}{2\pi} \partial_x \phi_{\alpha I}$, with $\phi_{\alpha I}$
the chiral boson fields satisfying the commutation relations $[\phi_{R/L I}(x), \phi_{R/L J}(y)] = \pm i \frac{\pi}{m}
sgn(x-y)$. $T_{IJ} = |T_{IJ}| e^{i\theta_{IJ}}$ are back-scattering amplitudes between counter-propagating edge states
$LI$ and $RJ$ of the two layers. The velocity of the edge modes, $v_0$, is assumed for simplicity to be
the same for all of the edge states. The diagonal entries of $\lambda$, $\lambda_{II}^{\alpha \alpha}$ are
normalized to 1, while the off-diagonal elements of the symmetric matrix $\lambda$ parameterize the
density-density interaction between the different edge channels. Tunneling between copropagating edge
states is irrelevant and not included.

\begin{figure}
\includegraphics[width=2.5in]{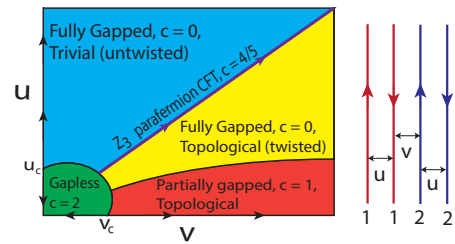}
\caption{\label{pdFig} Right: Counterpropagating edge states induced by gates. $1$ (red) and $2$ (blue) indicate
the layer index, and $u$, $v$ depict electron tunneling within and between layers, respectively.
Left: Expected phase diagram. 
$c$ indicates the central charge of the edge theory. In the absence of strong repulsive interactions,
$u$ and $v$ are irrelevant perturbations in the RG sense,
and only have a non-trivial effect if $u > u_c$ or $v > v_c$, where $u_c,~v_c$ is a proper ultraviolet
cutoff for the edge theory, which is typically on the order of the bulk energy gap.
A non-zero $T_{21} \propto u^2/E_{12}$ is induced to second-order
in perturbation theory, where $E_{12}$ is set by $v$ (see text),
so all counterpropagating edge states can be fully gapped in the limit
$|v| \gg |u|$, and $|v|, |u^2/E_{12}| > v_c$. The quantum phase transition between the two gapped
phases is the so-called $Z_3$ parafermion CFT. When the $u$, $\lambda$ perturbations become relevant
in the presence of strong repulsive interactions, then $v_c, u_c \rightarrow 0$, and the partially gapped
$c = 1$ region (red) shrinks to the $u  =0$ axis. }
\end{figure}

$\mathcal{H}_t$ contains competing pairs of cosine terms: the diagonal terms associated with $T_{11}$ and $T_{22}$
cannot be simultaneously minimized along with the off-diagonal terms associated $T_{12}$ and $T_{21}$, due to the
chiral nature of the edge states \cite{barkeshli2012a}. Consequently, this one-dimensional system
exhibits a rich phase diagram as a function of the interlayer tunneling.
For $m=3$, a qualitative phase diagram is displayed in Fig. \ref{pdFig}.
Here, we set $u \equiv T_{11} = T_{22}$ and $v \equiv T_{12}$, while $T_{21} = 0$. For now we consider
the tunneling amplitudes to be uniform in space. We have assumed the electron tunneling to be irrelevant,
so that the gapless chiral Luttinger liquid phase is stable for small $u,v$.

At small $u$ and large $v$, there is a partially gapped phase where two edges are coupled and gapped by $T_{12}$ and the other two edges remain gapless. At large $u$ and small $v$, there is a fully gapped phase where the FQH liquid in each layer is restored and the antidots disappear. Remarkably, even when $T_{21} = 0$, it is possible to obtain a ``twisted" fully gapped phase with coherent interlayer tunneling. This is because a non-zero $T_{21}$ will be generated perturbatively: $T_{21} \propto - u^2/E_{12}$, where $E_{12}$ is the energy gap of $(\phi_{L1} + \phi_{R2})$ that is generated by $T_{12}$. The perturbatively generated $T_{21}$ will then lead to an energy gap for
$(\phi_{L2} + \phi_{R1})$. With further increase of $u$, there will be a phase transition from this ``twisted" phase to the intra-layer dominant phase in which $\phi_{L1}+\phi_{R1}$ and $\phi_{L2}+\phi_{R2}$ are gapped. For $m=3$ the phase transition is described by a $Z_3$ parafermion conformal field theory \cite{barkeshli2013defect2,lecheminant2002,mong2013}. Taking into account the effect of disorder and the momentum
mismatch between edge states, $u$ and $v$ will be spatially random. The full phase diagram in the disordered case has
not yet been systematically studied, although the gapped and partially gapped phases of Fig. \ref{pdFig} will persist, with
the finite correlation length induced by the energy gap being replaced by a suitable localization length.

\it Topological degeneracy and the tunneling problem \rm -- To start with, we consider the TLJ formed
by the antidot edge states, while the tunneling with the outer edge is switched off. The top and bottom layers
are effectively glued to each other along the cut induced by the gates, so that the system obtains a
nontrivial genus, resulting in topological ground state degeneracies \cite{barkeshli2012a}. Each TLJ in a bilayer
$1/m$ state leads to $m$ topologically degenerate states, which can be labeled by the topological charge
$Q$ around a loop surrounding the TLJ. In each topological sector, there are edge quasiparticle excitations
both along the outer edge and along the antidot edge. The latter has an energy gap which is either
determined by finite size effect $\propto 1/l$ in the partially gapped phase, or by the induced gap
in the twisted gapped phase. For temperatures much lower than this gap, the low-energy Hilbert space
of the system can be written in the form
$\mathscr{H}_{loc;1} \otimes \mathscr{H}_{loc;2} \otimes \mathscr{T}$, where $\mathscr{H}_{loc;1} \otimes \mathscr{H}_{loc;2}$
describe local excitations on the outer edges, and $\mathscr{T}$ is a finite $m^n$ dimensional space that labels the topological sector,
for $n$ TLJs. The topologically distinct ground states have
an exponentially small splitting $\delta E \propto e^{- \ell/\xi_{loc}}$, where $\xi_{loc}$ is a suitable localization length associated with the interlayer backscattering that is responsible for creating the TLJ.
The physical processes that can mix the topologically distinct sectors correspond to quasiparticle
tunneling events across the entire TLJ, which we will discuss, along with $\xi_{loc}$ and the effect of
additional non-topological states in the TLJ, in the Supplemental Materials.

Now we consider tunneling between the outer edges across the QPC. The tunneling process of interest
is the quasiparticle tunneling from the outer edge of the top layer to the outer edge of the bottom layer via the TLJ.
Such a tunneling process can only occur when the QPC is in close proximity to one of the two end points of the TLJ.
At other points along the junction line, the edge states in one of the two layers is gapped, so that the tunneling
is suppressed. This is an observable consequence of the parafermion zero mode localized at each end point of the TLJ. 
The effective Hamiltonian can then be  written as $H(t) = \int dx [\mathcal{H}_0 + \mathcal{H}_b(t)] + \sum_{n=1}^m E_n |n\rangle \langle n |$, with $\left|n\right\rangle$ and $E_n$ labelling the topological ground states and their energies, and $\mathcal{H}_0 = m \pi v_0 \sum_{I,J} n_I \lambda_{IJ} n_J$ the Hamiltonian density of the outer edge. The tunneling Hamiltonian is 
\begin{align}
\label{ZnQIM}
\mathcal{H}_b(t) = \delta(x) \sum_{l=-\infty}^\infty  \lambda_l e^{i l e^* V_r t/ \hbar} e^{il  (\phi_{R1} - \phi_{R2})(t,x)} (T^+)^l
\end{align}
where $e^{il  (\phi_{R1} - \phi_{R2})(t,x)}$ is the quasiparticle hopping operator acting on the outer edges, and $T^+$ is the cyclic permutation operator acting on the topological ground states defined by $T^+ |n \rangle = |n + 1\rangle$ (with $|m \rangle \equiv |0\rangle$). $T^+$ satisfies $(T^+)^m = 1$. $e^* = e/m$ is the charge of the
quasiparticle that is tunneling from one edge to another, and $V_r(t)$ is the time-dependent potential
difference applied between the two layers.
Note that Hermiticity requires the coefficient $\lambda_l$ to satisfy $\lambda_l = \lambda_{-l}^*$. The case $m = 2$ can be re-fermionized
and mapped onto the problem of a Luttinger liquid lead coupled to a Majorana fermion zero mode \cite{fidkowski2012}, and
is closely related to an anisotropic Kondo impurity problem in the presence of a Zeeman field \cite{emery1992}.
The case $m > 2$ provides a novel cyclic $Z_m$ generalization of the Kondo impurity problem.

It is convenient to define $\mathcal{O}_l \equiv e^{il  (\phi_{R1} - \phi_{R2})(x=0)} (T^+)^l$, which leads
to a charge transfer of $le^*$ from the top edge to the bottom edge and shifts the state of the clock $l$ times. The tunneling current operator is
$I_t = \frac{ie^*}{\hbar} \sum_{l=-\infty}^\infty l e^{i l e^* V_r t/\hbar} \lambda_l \mathcal{O}_l$. The
tunneling current can be measured through the interlayer conductance along the outer edge.
In the absence of tunneling between the outer edges, the system exhibits a quantized
conductance: $I_r =  \frac{1}{m} \frac{e^2}{h} V_r$, where $V_r \equiv V_1 - V_2$ is the applied potential difference
and $I_r \equiv I_1 - I_2$ is the measured current difference between the layers. In the presence of the tunneling current,
there is a deviation from this quantized value:
$I_r= \frac{1}{m} \frac{e^2}{h} V_r - 2 I_t(V_r)$. We are interested in the nonlinear tunneling
conductance at finite voltages $V_r$ and temperature $T$: $G_t(V_r, T) = \frac{d I_t(V_r, T)}{dV_r}$.

{\it Weak tunneling limit and master equation approach.}-- In what follows, we will compute $G_t(V_r,T)$ in the weak tunneling limit, when the amplitude of the tunneling matrix element $\left|\lambda_l\right|$ is much smaller than the energy difference between the initial and final states. To obtain the quantitative criteria for weak tunneling, one needs to take into account the renormalization effect of the coupling constants $\lambda_l$. At an energy scale $\Lambda$, the renormalized coupling is $\lambda_l[\Lambda]=\left( \frac{\Lambda}{\Lambda_0} \right)^{\Delta_l - 1} \lambda_l^0$, with $\lambda_l^0$ the microscopic coupling constant at a high energy cutoff $\Lambda_0$. $\Delta_l$ is the scaling dimension of the tunneling operator $O_l$, which is $\Delta_l=\frac{l^2}{m}$ in the ideal edge theory (\it ie \rm no interactions between different edge states)\cite{wen1992}. Denoting $E_{ab}=E_a-E_b,~a,b=0,1,...,m-1$ as the energy difference between the clock states $\left|a\right\rangle$ and $\left|b\right\rangle$, the tunneling is a weak perturbation to the quantum clock degrees of freedom if $\lambda_l^0$ is small enough such that $\lambda_l [ |E_{n,n+l}-V_r| ] \ll max(|E_{n,n+l}-V_r|,T)$ for all $n$ and $l$.

In the weak tunneling limit, we can use a master equation approach, which uses the sequential tunneling approximation and assumes that different tunneling events are not coherent \cite{chamon1993,furusaki1998}. We let $p_n$ be the probability that the system is in the topological ground state $\left|n\right\rangle$, and $\Gamma_{n,n'}^l$ the rate of going from state $|n \rangle$ to $|n'\rangle$, with a charge transfer of $e[(n' - n)/m + l]$. These
will be determined below.
 Furthermore, let us define the net transition rate from $|n\rangle$ to $|n'\rangle$ as
$\Gamma_{n,n'} = \sum_{l =-\infty}^\infty \Gamma_{n,n'}^l + 1/\tau_{n,n'}$,
which in addition to a sum over $\Gamma_{nn'}^l$ also includes 
a temperature-dependent relaxation rate $1/\tau_{n,n'}$, which describes the relaxation due to bulk quasiparticle tunneling processes, without transferring charge between the outer edges. In thermal equilibrium, the density of quasiparticles is $\propto e^{-\Delta/T}$, and therefore $1/\tau_{nn'} \propto e^{\Delta/T}$. In the following we will take the simple approximation $1/\tau_{nn'}=\delta_{n,n'\pm 1}/\tau$. A simple estimate can relate $1/\tau$ to the longitudinal conductivity \cite{bonderson2009}: $1/\tau \sim \sigma_{xx} \Delta/e^2$, where $\Delta$ is the bulk energy gap.

For a static (DC) voltage $V_r$, the probabilities in the stationary state are determined by the master equation:
\begin{eqnarray}
\frac{d p_n}{d t} = \sum_{l=1}^{m-1} [\Gamma_{n+l,n} p_{n+l} - \Gamma_{n,n+l} p_n]=0,\label{mastereq}
\end{eqnarray}
The tunneling current is determined by a probability average $I_t = \sum_{n=0}^{m-1} p_n I_n$, where $I_n$ is the current in state $|n \rangle$ of the system:
$I_n = \sum_{l=-\infty}^\infty \sum_{a = 0}^{m-1} e(\frac{a}{m} + l )\Gamma_{n,n+a}^l$. To lowest order, the rates are:
\begin{align}
\Gamma_{n,n+l}^0 
&= \frac{|\lambda_l|^2}{\hbar^2}\int_{-\infty}^\infty e^{-i (l e^* V_r - E_{n,n+l})t'} [G^{(l)}(t')]^2 dt',
\end{align}
In our notation, $\Gamma_{n,n'}^l \equiv \Gamma_{n,n'+ml}^0$, where the state $|n'+ml\rangle \equiv |n'\rangle$.
Using the Green's function
$G^{(l)}(t) = \langle e^{i l \phi_R(t)} e^{-i l \phi_R(0)} \rangle$ $=e^{-i\frac{\pi l^2}{2m} sgn(t)} \left(\frac{ (\pi T/\Lambda_0)}{\sinh[\pi T |t|]}\right)^{l^2/m}$,
we obtain the explicit form
$\Gamma_{n,n+l}^0 = \frac{|\lambda_l|^2}{\hbar \Lambda_0} \left( \frac{\pi T}{\Lambda_0} \right)^{\frac{2 l^2}{m} - 1} B\left( \frac{l^2}{m} -i \bar{\Omega}, \frac{l^2}{m} + i \bar{\Omega} \right)
e^{-\pi \bar{\Omega}}$
where $\bar{\Omega} =  (l e^* V_r - E_{n,n+l})/2\pi T$ and $B(x,y)$ is the Beta function.

The equations above can be generalized to the case of an AC voltage difference,
$V_r(t) = V_{r0} \cos(\omega_0 t)$, by modifying the master equation:
$\pm i \omega_0 \bar{p}_{n;\pm \omega_0} = \sum_{l=1}^{m-1} [\Gamma_{n+l,n} \bar{p}_{n+l;\pm \omega_0} - \Gamma_{n,n+l} \bar{p}_{n;\pm \omega_0}]$,
where $\bar{p}_{n;\omega}$ is the Fourier transform of $p_n(t)$.

To understand the results of the master equation, let us first consider the case $m  = 3$, with
$\omega_0 = 1/\tau = 0$, and consider only single quasiparticle tunneling
$\lambda_1=\lambda,~\lambda_l=0$ for $|l| > 1$. In this case, a steady state current can only be
achieved if there is enough energy to surmount the energy barrier of adding one quasiparticle or quasihole for {\it every} state $|n\rangle$.
This requires the condition $e^* V_r > \max_{n=0,1,2}E_{n+1,n}$ or $-e^* V_r <\min_{n=0,1,2}E_{n,n+1}$. 
As a result, there are only two resonance peaks at $e^* V_r =\max_{n=0,1,2}E_{n+1,n}$
and $-e^* V_r =\min_{n=0,1,2}E_{n,n+1}$. Note that in the absence of fine-tuning of the energy splittings 
$E_{ab}$, the position of these two resonances are \it necessarily \rm asymmetric about zero energy. This is in direct
contrast to the case of Majorana zero modes, where there are two states, and the resonances are
symmetric.

If the relaxation rate $1/\tau$, or the frequency $\omega_0$, become larger than the
transition rates $\Gamma_{n,n + 1}$, then additional resonances appear when $e^* V_r = E_{n+1,n}$,
leading to 3 resonances in the case $m = 3$, as is shown in Fig. \ref{condFig}. In the DC limit $\omega_0 = 0$, the strength of the third peak provides a direct measure of the relaxation rate $1/\tau$,
relative to the transition rates of the system. Note that $1/\tau$ depends exponentially on $T$ and can be tuned strongly. More generally in a bilayer $1/m$ state, at strong enough $1/\tau$ or frequency $\omega_0$, the number of resonance peaks is equal to the number of topological ground states $m$ if we only include single quasiparticle tunneling. 
If we further consider multi-quasiparticle tunneling processes with nonzero $\lambda_l,~|l|>1$, more resonance peaks generically appear at energy values $l e^*V_r=E_{a,a+l}$.

In the Supplemental Materials we discuss (1) the alternative electron-hole bilayer setup, (2) the application to detecting
parafermion zero modes from superconductor - quantum Hall heterostructures, (3) additional contributions
to the interlayer conductance, (3) additional details about the topological degeneracy and splitting, (4) additional subtleties
about mapping to the cyclic quantum impurity model and effect of additional states in the TLJ, and (5) detection of higher topological degeneracies from higher genus
surfaces. 

\it Acknowledgments --\rm  We thank Moty Heiblum, J. Shabani, S. Das Sarma, R. Lutchyn, M. Cheng,
P. Bonderson, M.P.A Fisher, A. Ludwig, L. Balents, and Steven Kivelson for helpful discussions.
This work was partially supported by the Simons Foundation (MB) , the David and Lucile Packard Foundation (XLQ), and 
by WIS-TAMU, ISF, and ERC (FP7/2007-2013) 340210 grants (YO).


\newpage

\appendix

\newpage

\begin{widetext}

\section{Supplemental Materials}

\subsection{I. Second experimental setup: electron-hole bilayers}

In this section we briefly introduce a second experimental platform where topological ground state degeneracy
can be probed. The setup is shown in Fig. \ref{ehBilayer}, and consists of an electron-hole bilayer system where one layer
consists of electrons while the other consists of holes, at the same density. Such electron-hole bilayers have
been investigated experimentally in recent years \cite{seamons2007} . In the presence of a perpendicular magnetic field, and
for large enough interlayer separation, the electron layer can form a $\nu = 1/m$ Laughlin FQH state while
the hole layer forms a $\nu = -1/m$ Laughlin FQH state. Next, we form antidots in the two layers using top and bottom gates,
such that the antidots in the two layers are directly on top of each other. This is in contrast to the setup of the main text,
where the antidots are offset laterally.

The edge states propagating around the antidots are propagating in different directions, due to the fact that
one layer consists of holes and the other electrons. Therefore, sufficiently large vertical electron
tunneling can generate an energy gap between the counterpropagating antidot edge states of the different layers.
Under this condition, the top layer is effectively ``glued'' to the bottom layer, in the sense that a quasiparticle
from the top layer can coherently propagate into the bottom layer at low energies. The pair of antidots can be thought
of as giving rise to a ``wormhole'' that connects the top and bottom layers, and the system can now be viewed
as a single FQH state on a higher genus surface. For a single wormhole, the system is topologically equivalent
to a single $1/m$ Laughlin FQH state on a cylinder, which has $m$ topologically distinct sectors.

\begin{figure}
\includegraphics[width=2.5in]{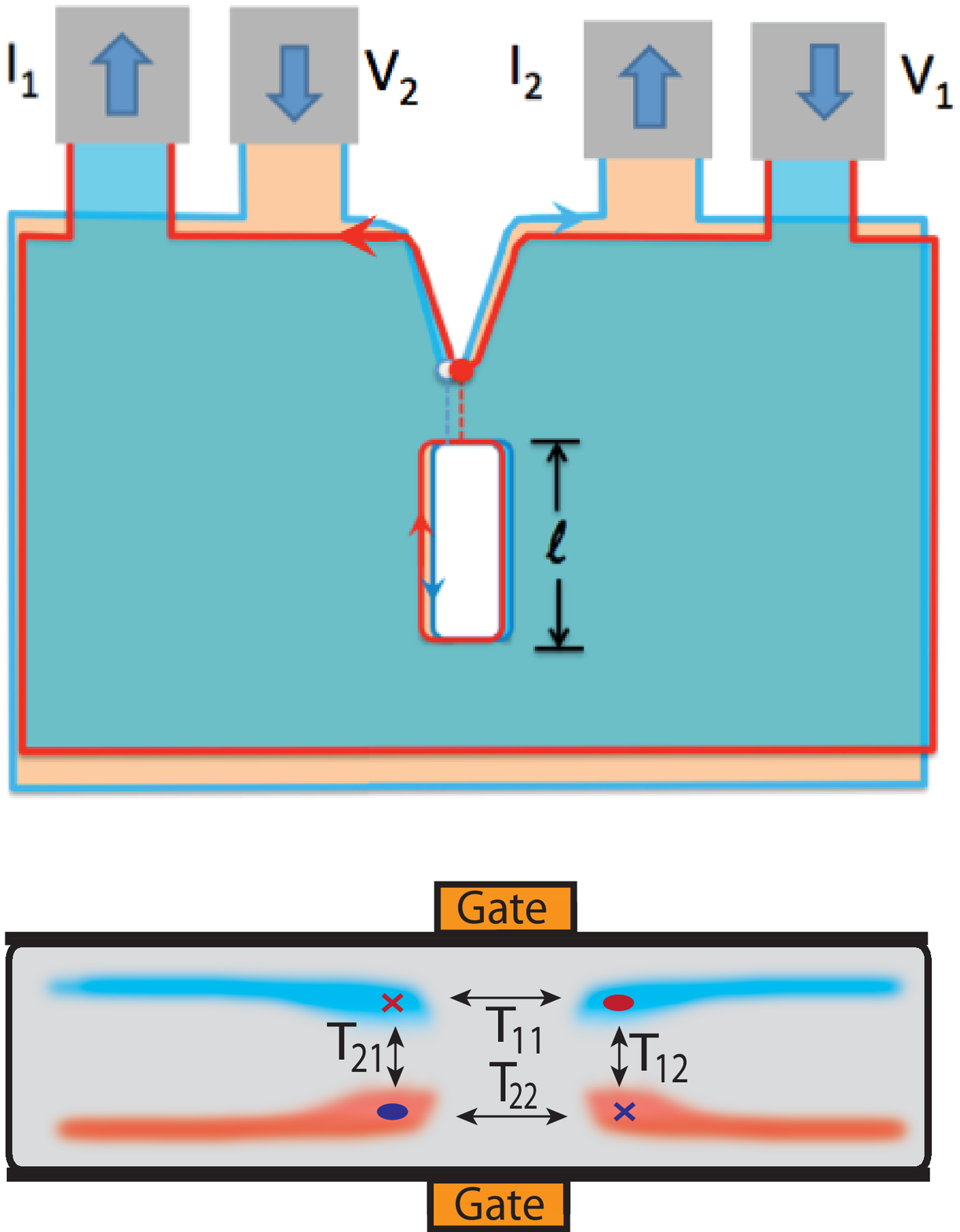}
\caption{\label{ehBilayer} Alternative experimental setup using electron-hole bilayers. One
layer forms a $\nu=1/m$ Laughlin FQH state, while the other layer forms a $\nu=-1/m$ Laughlin
FQH state. The top and bottom gates are aligned to be directly on top of each other. The edge states
of the antidots propagate in different directions in the two layers. }
\end{figure}

Many of the considerations of the main text can be directly applied to this setup as well. The topological
degeneracy can be directly probed by weakly splitting the states through the finite size effect.
Quasiparticles can tunnel from the outer edge of one layer to the outer edge of the other layer through the wormhole, and
therefore the QPC geometry of Fig. \ref{ehBilayer} can directly be mapped onto the $Z_m$ clock impurity
model studied in the main text.

There is, however, one crucial difference as compared to the setup of the main text. Recall that the setup in the
main text possesses $Z_m$ parafermion zero modes at the ends of the TLJ \cite{barkeshli2012a,barkeshli2013genon,
barkeshli2013}. The physical consequence of this is that at the location of the parafermion zero modes,
excitons (quasiparticle-quasihole pairs from different layers) can be absorbed or emitted.
In the current electron-hole setup, there is no special point where quasiparticle-quasihole pairs from the
different layers (excitons) can annihilate (this can happen anywhere along the line junction),
and there are therefore no localized zero modes. Furthermore,
in \cite{barkeshli2013genon} it was shown that an adiabatic $2\pi$ rotation of the TLJ realizes a non-trivial braiding operation
of the zero modes, leading to non-Abelian statistics. In contrast, in the electron-hole bilayer, an adiabatic $2\pi$ rotation
of the wormhole has no effect on the topological ground states.

Nevertheless, despite these differences, this system can also be used to probe topological ground state degeneracies,
and may have its own advantages as compared with the electron-electron bilayers discussed in the main text.

\subsection{II. Application to parafermion zero modes in single-layer FQH line junctions proximited with superconductivity}

In \cite{lindner2012,clarke2013,cheng2012,mong2013,oreg2013}, it was shown that 
parafermion zero modes can be realized in a somewhat different physical context. 
Consider a line junction in a single-layer FQH state, which possesses chiral FQH edge states on each side of
the junction. The edge states can be gapped in two ways: either (1) through normal backscattering, 
which will heal the FQH fluid on either side of the junction into a uniform FQH fluid, or (2) through 
Cooper pairing terms that are induced through proximity effect from a nearby
superconductor. In the second case, the edge states of the line junction enter into a gapped topological phase with respect to the surrounding quantum Hall fluid. The domain wall between the topological phase of the edge states, 
induced by the superconducting proximity effect,
and the normal gapped phase, induced by normal backscattering, will localize exotic parafermion zero modes. These parafermion zero
modes have a quantum dimension $\sqrt{N}$, so that each pair together leads to $N$ topologically degenerate states. As discussed in
\cite{barkeshli2013genon}, these are similar to the parafermion zero modes found in the bilayer FQH setups.

As in the setup described in the main text, the topologically degenerate states can be weakly split by tuning the size of the topological region
that is gapped by the superconducting proximity effect. Furthermore, a single quasiparticle can tunnel onto the zero mode at zero energy,
and such a process changes the topological state. Therefore, the geometry shown in Fig. \ref{scfqh} can be described by the following Hamiltonian:
\begin{align}
H = \int dx \mathcal{H}_{0;outer} + H_t + \sum_n E_n |n \rangle \langle n |,
\end{align}
where
\begin{align}
H_t = \sum_{l=-\infty}^\infty \lambda_l e^{i l e^* V t/\hbar \phi(x = 0, t)} (T^+)^l,
\end{align}
and $\mathcal{H}_{0;outer}$ is the kinetic term for the Hamiltonian density of the outer edge. This is equivalent to the
$Z_N$ quantum impurity model presented in the main text, where $\phi$ takes the place of $\phi_{R1} - \phi_{R2}$
in the bilayer system.

In the setup of Fig. \ref{scfqh}, the conductance of the outer edge will be given by
\begin{align}
G \equiv \frac{dI}{dV} = \nu \frac{e^2}{h} - G_t,
\end{align}
where $G_t \equiv \frac{dI_t}{dV}$ is the tunneling conductance for quasiparticles tunneling into the zero modes. $G_t$ will therefore
show resonances at finite voltage associated with the number of different topological sectors and their energy splittings,
providing a direct probe into the topological states generated by the $Z_N$ parafermion zero modes.

\begin{figure}
\includegraphics[width=2.5in]{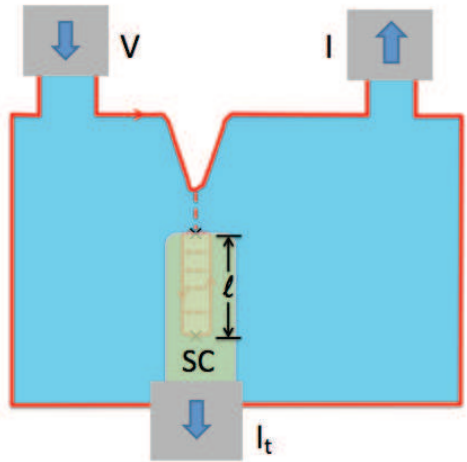}
\caption{\label{scfqh}  A line junction in a single layer FQH state gives rise to a chiral Luttinger liquid on each side of the junction.
Superconducting proximity effect can localize the edge states, in such a way that the ends of the line junction localize parafermion
zero modes, giving rise to topological degeneracies. Quasiparticle tunneling from the outer edge onto the zero modes shifts
the system among the different topological sectors. Therefore, weakly splitting the topological degeneracy leads to
finite voltage resonances in the conductance measured from the outer edge. }
\end{figure}

\subsection{III. Additional contributions to interlayer conductance}

Here we will briefly discuss additional effects that have important implications for the
interlayer conductance $dI_r/dV_r$ of the outer edges in the bilayer setup of the main text.
These include:

\begin{enumerate}
\item Interlayer electron tunneling along the edge.

While electron tunneling between the two edges will provide a non-zero contribution to the interlayer
conductance, this contribution will be a smooth function of the voltage difference applied between
the two layers, due to the smooth electron density of states at the edge. Furthermore, by defining the
edge of the sample with gates, it is possible to spatially separate the outer edges of the two layers, thus allowing
for control of interlayer tunneling along the outer edge.

\item Impurity states in the bulk, away from the TLJ, that may lead to stray resonances in the interlayer conductance.

Another way for current to flow from one edge to another at low energies below the bulk gap is through
impurity states in the bulk. If there are two impurity states in each of the two layers, with the possibility
of electron tunneling between them, then it is possible that to have a process where
electrons tunneling from the edge onto the impurity state in the top layer, then onto the impurity state in the
bottom later, and finally onto the bottom edge. Such a process is expected to be strongly suppressed,
however its existence is possible in principle, and it may lead to finite voltage resonances in the interlayer
tunneling conductance. These contributions can be distinguished from resonances due to the topological states
in several ways. The easiest way is to compare the conductance with that at filling fraction $\nu = 1$ in each layer,
where there are no topological degeneracies. The energy levels of the impurity states will presumably be independent of the magnetic
field, and therefore any spurious resonances will also be present at $\nu =1$. Direct comparison can then
help distinguish resonances from topological states from other spurious resonances.
Furthermore, the number of resonances from the topological states are universal and
independent of sample details, while other resonances depend on details of the impurity
configuration. Therefore systematic studies of tunneling into different sides of the TLJ, and in
different samples, can be used to distinguish the stable resonances due to topological states
from any spurious resonances due to impurity states.

\item Effect of additional states in the TLJ.

We will comment on this in the subsequent section on mapping to the $Z_N$ quantum impurity model.

\end{enumerate}

\subsection{IV. Splitting of topologically degenerate states: Additional discussion}

Here we will discuss the splitting of the topologically degenerate states in some more detail.

\subsubsection{Mapping to squeezed cylinder}

\begin{figure}
\includegraphics[width=4in]{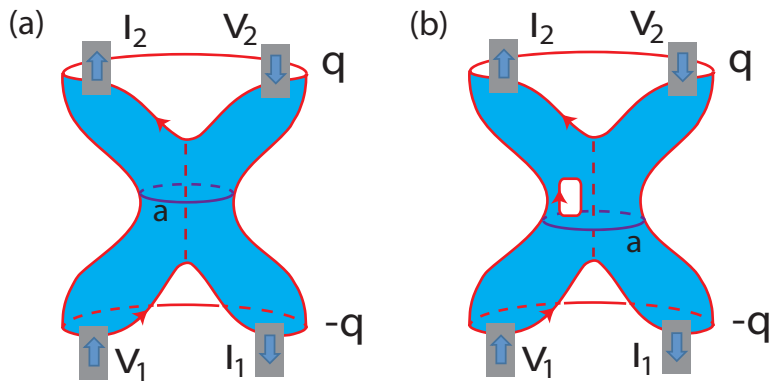}
\caption{\label{squeezedCyl} The setup of Fig. \ref{setupFig} of the main text can be mapped to a single layer FQH state on the surface of a
squeezed cylinder, shown here. (a) The case where the TLJ is in the fully gapped topological (twisted) phase. (b) The case where the TLJ
is in the partially gapped topological phase (red region in Fig. \ref{pdFig} of main text). The ungapped edges of the TLJ are mapped to a hole
on the surface of the cylinder. $(q,-q)$ labels the number of units of fractional charge, modulo $m$ for a $1/m$ Laughlin state, on the two edges. 
Squeezing the middle of the cylinder enhances quasiparticle tunneling around the loop $a$, which gives an energy splitting $\propto e^{-\alpha \ell/\xi}$ to the different
sectors, where $\ell$ is the length of the loop $a$, $\alpha$ is a constant of order one, and $\xi$ is the bulk correlation length.
The physics of a quantum point contact (QPC) connecting the two edges can then be modelled by the $Z_N$ quantum impurity model introduced
in the main text: the splitting leads to finite voltage resonances in the tunneling conductance of the QPC.
}
\end{figure}

Let us consider the setup of Fig. \ref{setupFig} in the main text, where there is a single TLJ in the bilayer geometry. Topologically,
this geometry is equivalent to the squeezed cylinder shown in Fig. \ref{squeezedCyl} , and it is clarifying to understand the
physics in this simpler geometry. In the $1/3$ Laughlin state, there are 3 topologically distinct sectors associated with the fractional charge (modulo $1$)
on each edge: $(0,0)$, $(1/3,-1/3)$, and $(-1/3, 1/3)$. In the limit that the top and bottom edges are infinitely long, these
different topological sectors are degenerate. However, squeezing the cylinder allows instanton processes where
a quasiparticle-quasihole pair is created out of the vacuum, one of them travels around the loop $a$ shown in Fig. \ref{squeezedCyl},
and they reannihilate. This process, which we denote as $W(a)$, can measure the topological sector and therefore will
split the degeneracy between topological sectors. The amplitude for this process is exponentially suppressed in the ratio
$\ell/\xi$, where $\ell$ is the length of the loop $a$ and $\xi$ is the correlation length of the FQH system. Therefore,
decreasing $\ell/\xi$ increases the splitting between the topological states.

Quasiparticle tunneling between the outer edges of the cylinder shifts the topological sectors. Therefore, the interlayer
conductance $dI_r/dV_r$, where $I_r \equiv I_1 - I_2$ and $V_r \equiv V_1 - V_2$, will be given by the $Z_N$ quantum
impurity model presented in the main text.

Note that in the setup of Fig. \ref{setupFig}a, it is possible that the edge states created by the top and bottom gates are
either partially gapped ($c= 1$ partially gapped topological phase in Fig. \ref{pdFig} of main text), or fully gapped ($c = 0$ fully gapped
topological phase in Fig.\ref{pdFig} of main text). In the mapping to the squeezed cylinder,
for the former case the edge states flowing around the gates map to a hole on the surface of the cylinder (Fig. \ref{squeezedCyl}b). In the latter
case, we obtain Fig. \ref{squeezedCyl}a. 

\subsubsection{Physical origin of splitting: bulk vs. line junction processes}

Let us consider the TLJ induced in the double layer system. The system has three important lengths scales: $\xi_{loc;1}$, $\xi_{loc;2}$,
and $\xi_{bulk}$. $\xi_{bulk} \propto 1/\Delta_B$, where $\Delta_B$ is the bulk energy gap, is the correlation length of the bulk
FQH fluid. In the presence of interlayer tunneling, the boson mode $(\phi_{L1} + \phi_{R2})$ is localized, with localization length $\xi_{loc;1}$,
and $(\phi_{L2} + \phi_{R1})$ is localized with a different localization length, $\xi_{loc;2}$. If only one pair of counterpropagating
edge states are localized (partially gapped twisted phase in Fig. \ref{pdFig} of main text),
the system is still in the topological phase and will have topological degeneracies, but
we will have $\xi_{loc;2} \rightarrow \infty$.

The question of the physical origin of the splitting of the topological sectors depends on the relative magnitudes
of min($\xi_{loc;1}$, $\xi_{loc;2}$), and of $\xi_{bulk}$. There are two different quasiparticle tunneling processes that
will split the three topological sectors induced by a single TLJ in the bilayer system with the $1/3$ Laughlin state in each
layer. One of them is a quasiparticle-quasihole being created in a given layer in the bulk, one of them propagating around the
gate, and then reannihilating. This process is analogous to the one described above in the context of the squeezed cylinder,
around the loop $a$ in Fig. \ref{squeezedCyl}. The amplitude for this process is $\propto e^{- \alpha \ell/\xi_{bulk}}$, where $\ell$
is the length of the gates, and $\alpha$ is a constant of order one.

The other process that can measure the topological sector is a quasiparticle-quasihole pair from different layers (\it ie \rm
a fractional exciton), being emitted from one zero mode, propagating through the edge states, and being absorbed at the other.
The amplitude for this process is $\propto e^{- \ell/\xi_{loc}}$, where $\xi_{loc} = min(\xi_{loc;1}, \xi_{loc;2})$,
because the smaller localization length will limit the strength of this process.

Which of these physical processes is the one responsible for splitting the states depends on whether
$\xi_{bulk} \gg min(\xi_{loc;1}, \xi_{loc;2})$, in which case the bulk process is more favorable, or
$\xi_{bulk} \ll min(\xi_{loc;1}, \xi_{loc;2})$, in which case the exciton process through the TLJ is more favorable.

\subsection{V. Mapping to $Z_m$ quantum impurity model: Additional discussion}

In the main text, we argued that in the geometry of Fig. \ref{setupFig}, the interlayer conductance can be understood
by mapping the low energy theory to a $Z_m$ quantum impurity model. Here, the only low energy states coming from
the TLJ were considered to be the topological degenerate ground states, while the rest of the states were associated
with local excitations on the outer edge. However, it is indeed possible that the TLJ contributes additional low energy
states, not included explicitly in the $Z_m$ quantum impurity model. Here we would like to briefly analyze the effect of these
additional possible states. There are several different possible origins of additional low-lying states, which we will
analyze in turn:
\begin{enumerate}
\item Localized states due to disorder in the TLJ.

\item Additional gapless states due to edge reconstruction.

\item The edge states induced by the gates are in the fully gapped topological phase (see Fig. \ref{pdFig}), with a gap
$\Delta_{edge} < \Delta_B$, where $\Delta_B$ is the bulk energy gap.

\item The edge states induced by the gate are partially gapped (see Fig. \ref{pdFig}) but in the topological phase; that is
the combination $(\phi_{L1} + \phi_{R2})$ is gapped, but $(\phi_{L2} + \phi_{R1})$ is gapless.

\item Either of the previous two situation, but now where $(\phi_{L2} + \phi_{R1})$ acquires a finite size gap ($\propto 1/\ell$)
due to the finite size $\ell$ of the TLJ.
\end{enumerate}
In (1), (3), and (5), we expect that the interlayer conductance will show satellite peaks, associated with the possibility of quasiparticle
tunneling shifting the energy of the clock, while exciting nearby states in energy. In cases (2) and (4), we expect that the possibility
of exchanging energy with the gapless modes of $(\phi_{L2} + \phi_{R1})$ while tunneling the quasiparticles will broaden the
resonances. The effects of this can be modelled by including another edge state, and including terms that couple to this edge state.
While this may modify quantitative features of the resonance line shape, we do not expect that it will alter the resonances in any
qualitative way.

\subsection{VI. Higher genus surfaces and higher topological degeneracies }

As explained in \cite{barkeshli2013}, every pair of gates increases the effective genus of the surface by $1$. Therefore,
in a double layer system with a $1/3$ Laughlin FQH state in each layer and in the presence of $n$ TLJs, the system will
have $3^n$ topologically distinct sectors, which are associated with the $2n$ $Z_3$ parafermion zero modes localized at the ends of each TLJ.
As explained in the main text, the Hilbert space of the system can be written as
$\mathscr{H}_{loc;1} \otimes \mathscr{H}_{loc;2} \otimes \mathscr{T}$, where $\mathscr{H}_{loc;1} \otimes \mathscr{H}_{loc;2}$
describe local excitations on the outer edges, and $\mathscr{T}$ is a finite $3^n$ dimensional space that describes the topological sectors.

The $Z_3$ localized parafermion zero modes can be described by operators $\alpha_i$, where a one-dimensional ordering of
the zero modes must be chosen. The algebra of these operators can be written as:
\begin{align}
\alpha_i \alpha_j = \alpha_j \alpha_i e^{2\pi i sgn(i - j)/3}.
\end{align}
Physically, $\alpha_i$ corresponds to the creation operator for the fractional exciton at the ends of the TLJs. However a single
operator $\alpha_i$ by itself is unphysical: Fractional quasiparticles cannot simply be created out of the vacuum, as the total topological
charge of the system should be trivial. One class of physical operators are the bilinears $\alpha_i^\dagger \alpha_j$.

In the presence of a finite separation between the zero modes, the fractional excitons can be emitted/absorbed at different
zero modes, which leads to an effective Hamiltonian for the topological subspace:
\begin{align}
H_{top} = \sum_{ij} A_{ij} \alpha_i^\dagger \alpha_j + H.c.
\end{align}
The tunneling Hamiltonian from the outer edge onto one of the zero modes becomes:
\begin{align}
H_t = \sum_{l = -\infty}^\infty \lambda_l e^{i l e^* V_r t / \hbar} e^{i l (\phi_{R1} - \phi_{R2})(x=0)} (\alpha_1)^l
\end{align}
Now we can study the interlayer tunneling conductance for any set of couplings $A_{ij}$ between the zero modes, using the
methods developed in this paper.

For generic couplings $A_{ij}$, all $3^n$ states will be split, and can be directly probed through the interlayer conductance,
as shown in Fig. \ref{genus2Fig}.

\begin{figure}
\includegraphics[width=2.5in]{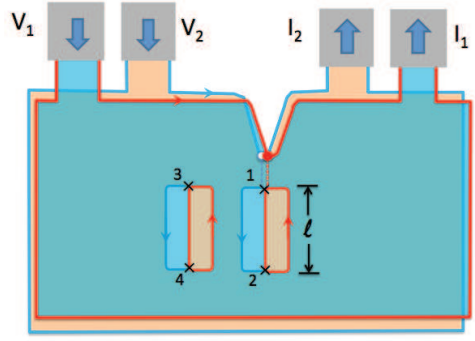}
\caption{\label{genus2Fig} In the presence of additional TLJs, the topological degeneracy increases exponentially. These topological
degeneracies are split by quasiparticle tunneling processes when the size of and distance between TLJs is finite. Tunneling of quasiparticles
from the edge can probe the different topological states, as shown.  }
\end{figure}

\end{widetext}

\end{document}